\documentclass[cameraready]{Interspeech}
\usepackage{multirow}
\usepackage{subfig}
\usepackage{hyperref}
\title{Sleep Sound Event Detection Powered by Learnable Multi-Resolution Adaptive Line Enhancer}

\author[affiliation={1}]{Chanwoo}{Park}
\author[affiliation={1}, correspondingauthor]{Chanwoo}{Kim}

\address{
$^1$ Department of Artificial Intelligence, Korea University, South Korea
}

\email{\{cksdn1290, chanwcom\}@korea.ac.kr
}


\usepackage{comment}

\begin{document}

\maketitle
\footnotetext{The code is available at \url{https://github.com/honeysleep/sed}}
\begin{abstract}
Sleep-disordered breathing, including snoring, hypopnea, and obstructive apnea, is the hallmark of obstructive sleep apnea, whose severity is quantified by the Apnea-Hypopnea Index (AHI).
As conventional polysomnography is costly and clinically demanding, microphone-only detection has emerged as a noninvasive alternative for screening.
We propose ALE-Confidence Fusion Sound Event Detection (ACF-SED), the first framework that directly injects the per-frame confidence map of an Adaptive Line Enhancer (ALE) into Transformer attention biasing and feature modulation, rather than as mere preprocessing.
A Multi-Resolution ALE Bank aggregates three parallel NLMS filters at distinct decorrelation delays, while a Confidence-Guided Cross-Path block couples event and noise streams via gated cross-attention.
On the Audio-Polygraphy Dataset for Sleep Apnea Analysis, ACF-SED achieves state-of-the-art detection across Event-F1, Segment-F1, and PSDS, enabling end-to-end AHI estimation.
\end{abstract}

\section{Introduction}
Sleep is an essential physiological process for maintaining both physical and mental health.
Insufficient sleep or poor sleep quality is closely associated with various health problems, including cardiovascular disease, cerebrovascular disease, stroke, and depression \cite{iannella:2022:obstructive,zhao:2023:association}.
Among these, obstructive sleep apnea (OSA) is the most common sleep disorder across all age groups.
According to the World Health Organization (WHO) \cite{bousquet:2007:global}, OSA is defined as a chronic condition characterized by repeated partial (hypopnea) or complete (apnea) interruptions of breathing during sleep caused by upper airway obstruction.

The severity of obstructive sleep apnea (OSA) \cite{strollo:1996:obstructive} is typically assessed using the Apnea–Hypopnea Index (AHI), which represents the number of apnea and hypopnea events per hour of total sleep time.
According to the criteria of the American Academy of Sleep Medicine (AASM) \cite{thornton:2012:aasm}, an AHI of less than 5 is considered normal, 5 to less than 15 indicates mild OSA, 15 to less than 30 indicates moderate OSA, and an AHI of 30 or higher indicates severe OSA.
The current gold standard for diagnosing OSA is polysomnography (PSG), a comprehensive test conducted in sleep centers in which multiple sensors are attached to the patient’s body to monitor various physiological signals overnight.
However, PSG has several limitations, including high cost, long waiting times, patient discomfort, and the need for specialized facilities and trained personnel, making it impractical for large-scale population screening \cite{pordzik:2024:waiting}.
A simplified alternative, polygraphy (PG), is also widely used, but because it does not include electroencephalography (EEG) to determine actual sleep time, it carries the risk of underdiagnosing AHI \cite{tan:2014:overnight}.

Due to the limitations of these conventional diagnostic methods, OSA is often underdiagnosed, in part because access to standard diagnostic procedures is limited.
This challenge is particularly pronounced among patients living in remote areas or regions with insufficient medical services, where access to such testing is even more restricted.
Consequently, there is a clear and pressing need for the development of noninvasive, low-cost, and robust methods for OSA detection and assessment \cite{kapur:2017:clinical,epstein:2024:use}.
Patients with OSA often produce loud snoring sounds due to increased upper airway resistance caused by partial pharyngeal collapse, which is recognized as the most common symptom of OSA.
Recent studies have highlighted that distinguishing between snoring and non-snoring sounds can serve as an initial step in OSA analysis, leading to growing interest in audio-based approaches that rely on signal processing and artificial intelligence techniques \cite{cao:2024:automatic,kim:2025:sound}.

The significance of these events extends beyond OSA itself.
Sleep-disordered breathing is increasingly recognized as a risk factor and early biomarker for cognitive decline and dementia \cite{yaffe:2011:sleep,bubu:2020:obstructive}.
This has motivated multimodal sleep analysis that models audio together with electroencephalography (EEG) and other signals \cite{phan:2022:sleeptransformer,park2025machine}.
More recently, large language models (LLMs) have fused such heterogeneous streams for neurocognitive screening, including dementia detection \cite{Park2025,park25d_interspeech}.
Since EEG-based pipelines require electrodes and specialized facilities, a robust audio-only detector remains valuable as a low-cost screening tool and acoustic front-end for such systems.

In this paper, we propose ACF-SED (ALE-Confidence Fusion-Sound Event Detection), a framework that directly injects the per-frame ALE confidence map derived solely from the adaptive filtering process into the attention and feature-modulation mechanisms for sleep sound event detection.
Overall, our contributions are as follows:

\begin{itemize}
  \item We propose ALE-Confidence Fusion Sound Event Detection (ACF-SED), which directly injects ALE confidence maps into Transformer attention biasing and into feature modulation, enabling the adaptive filter's internal state to be exploited within a deep learning framework.                                                                                                                        
  \item We introduce a multi-resolution ALE bank (MRAB) with three parallel NLMS filters at distinct decorrelation delays, and a Confidence-Guided Cross-Path Block (CCP) that asymmetrically processes event and noise streams via gated cross-attention.
  \item ACF-SED achieves \textbf{state-of-the-art performance} on the APSAA dataset \cite{gonzalez:2024:audio} across Event-F1, Segment-F1, and PSDS, supporting end-to-end AHI estimation from ambient audio.
\end{itemize}

\section{Related Works}
Sound event detection (SED) is a fundamental research area in audio signal processing that aims to identify the presence and temporal boundaries of sound events within audio recordings.
With the recent advancements in deep learning, the performance of SED systems has significantly improved, and models based on convolutional neural networks (CNN), recurrent neural networks (RNN) \cite{cakir:2017:convolutional}, and Transformers \cite{gong:2021:ast,naman:2025:fast,liu:2023:cat} have become dominant approaches in this field.

\subsection{Deep Learning Architectures for Sound Event}
Convolutional Recurrent Neural Networks (CRNN) \cite{cakir:2017:convolutional} have emerged as particularly effective architectures for polyphonic sound event detection.
The combination of CNNs for feature extraction and RNNs for temporal modeling has been shown to significantly outperform traditional approaches across various sound event detection datasets.
While the CNN layers extract high-level features invariant to local spectral and temporal variations, the RNN layers capture long-term temporal context within audio signals.
This hybrid approach has been widely adopted in subsequent studies and forms the foundation of many SED systems.

In addition to CRNNs, transformer-based architectures have recently gained significant attention in the audio domain.
The Audio Spectrogram Transformer (AST) \cite{gong:2021:ast} applies the Vision Transformer architecture to audio signals by converting them into spectrogram representations, achieving state-of-the-art results on multiple audio classification benchmarks such as AudioSet \cite{gemmeke:2017:audio}, ESC-50 \cite{piczak:2015:esc}, and Speech Commands V2 \cite{warden:2018:speech}.
The AST architecture completely removes the reliance on CNNs, demonstrating that a purely attention-based model can be sufficient for audio classification tasks.
More recently, efficient transformer variants such as Fast Audio Spectrogram Transformer (FAST) \cite{naman:2025:fast} and Causal Audio Transformer (CAT) \cite{liu:2023:cat} have been proposed, reducing computational complexity while maintaining competitive performance.

\subsection{Audio Feature Representation}
The choice of audio feature representation has a significant impact on the performance of SED systems.
Mel-spectrograms and mel-frequency cepstral coefficients (MFCC) are among the most commonly used features for audio signal analysis and classification \cite{wolf-monheim:2024:spectral}.
A mel-spectrogram applies a mel-scale filter bank to the power spectrogram, providing a time–frequency representation aligned with human auditory perception.

Recent studies \cite{turab:2022:investigating,wolf-monheim:2024:spectral} have compared the effectiveness of various audio features in deep learning-based classification tasks.
Comparative analyses indicate that Mel-spectrogram representations generally outperform MFCCs when used with CNN architectures, as they preserve more spatial information in the time–frequency domain.
Some studies have also explored combining multiple features to leverage their complementary strengths.

\subsection{Supervised and Self-Supervised Learning}
One of the main challenges in SED research is the scarcity of strongly labeled data with precise temporal annotations.
Weakly supervised sound event detection (WSSED), which relies only on clip-level labels without timestamp information, has emerged as a practical alternative \cite{deshmukh:2021:improving,xin:2023:background}.
In WSSED, a multiple instance learning (MIL) framework \cite{shah:2018:closer,miyazaki:2020:weakly} is commonly employed, where an audio clip is treated as a bag and its frames as instances.
Attention mechanisms and pooling strategies play a crucial role in aggregating frame-level predictions into clip-level outputs while preserving localization capability.

Self-supervised learning (SSL) approaches have also shown the potential to learn audio representations from unlabeled data.
These methods leverage pretext tasks to learn general-purpose features that can be fine-tuned for downstream applications.
Recent studies \cite{nguyen:2023:improving,yadav:2025:axlstms} have demonstrated that self-supervised pretraining on large-scale datasets such as AudioSet can significantly improve performance across various audio tasks.

\section{Proposed Method}

\subsection{Learnable Adaptive Line Enhancer (ALE)}
An adaptive filter \cite{farhang:2013:adaptive} is a system that has a linear filter with a transfer function controlled by variable parameters and a mechanism to adjust those parameters according to an optimization algorithm \cite{lan:2023:variational}.
Due to the complexity of the optimization process, almost all adaptive filters are implemented as digital filters.
In some applications, adaptive filters are required because certain parameters of the target processing task (for example, the positions of reflective surfaces in a reverberant space) are either unknown in advance or subject to change.
A closed-loop adaptive filter improves its transfer function using feedback in the form of an error signal \cite{wen:2024:adaptive}.

The term “adaptive” refers to the ability to adjust or respond to changing data or environmental conditions over time.
An adaptive filter, therefore, is one whose filter coefficients or tap weights change over time according to a specific criterion, depending on the characteristics of the input signal.
Unlike conventional filters with fixed coefficients, an adaptive filter optimizes its performance by continuously tuning its parameters to match environmental or signal variations \cite{comminiello:2018:adaptive}.

An Adaptive Line Enhancer (ALE) is a specific type of adaptive filter designed to extract narrowband signals buried in broadband noise.
Instead of requiring a separate reference signal, the ALE uses a delayed version of the input itself as a reference.
By doing so, it enhances periodic or slowly varying components (such as sinusoids) while suppressing uncorrelated noise components.
This makes ALEs particularly useful in applications such as speech enhancement, biomedical signal denoising, and detection of periodic signals in noisy environments \cite{vaseghi:2008:advanced,sathesh:2015:application,alipoor:2017:utilizing}.

\begin{figure}[htb!]
  \centering
  \includegraphics[width=\linewidth]{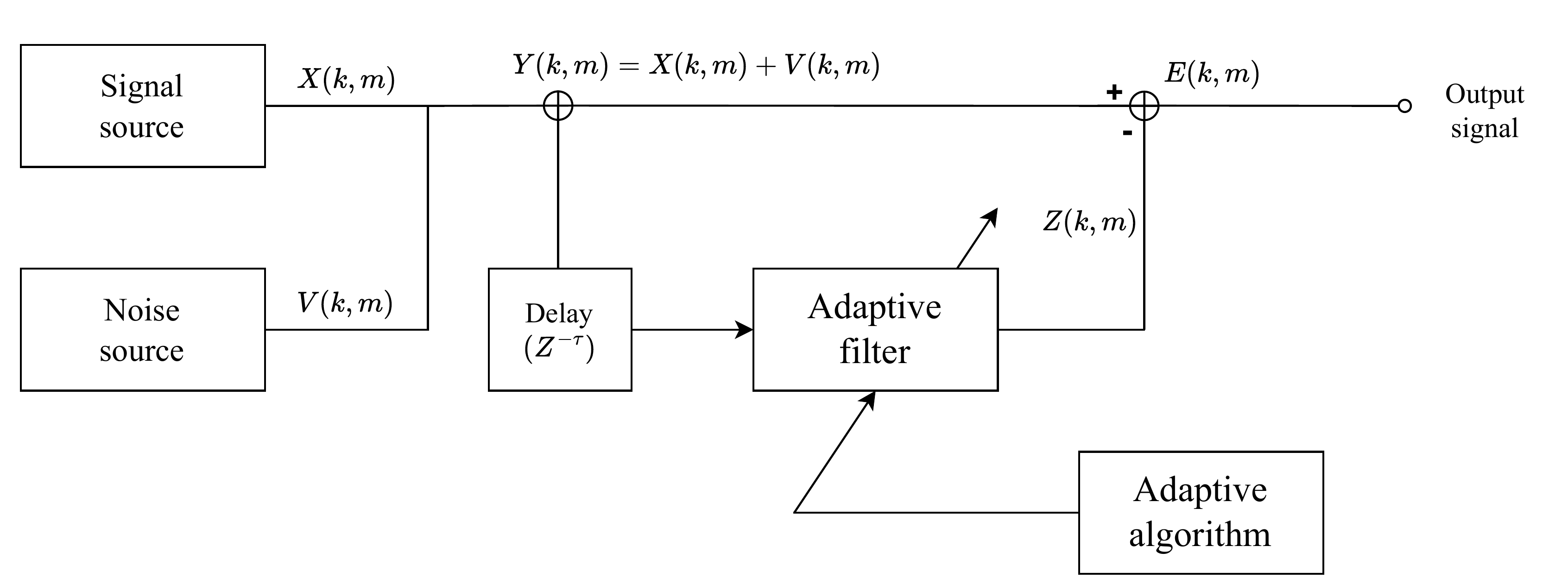}
  \caption{Block diagram of STFT domain adaptive line enhancer (ALE) system.}
  \label{fig:ale}
\end{figure}

The ALE is a specialized form of Adaptive Noise Canceling (ANC) that operates using only a single microphone.
The core principle of the ALE is to separate the periodic and stochastic components of a signal by employing a delay element.
The ALE can also be implemented in the short-time Fourier transform (STFT) domain, where the adaptive filter offers low computational cost and can be computed separately for each frequency bin \cite{nakanishi:2013:speech,atkins:2021:adaptive}.
This is illustrated in Figure~\ref{fig:ale}.
Specifically, a delayed version of the input signal is used as a reference.
The periodic components are extracted by the adaptive filter, usually using a normalized least-mean-square (NLMS) algorithm \cite{haykin:2002:adaptive}, operating as follows \cite{atkins:2021:adaptive}:

\begin{itemize}
  \item Periodic components: maintain correlation between the input and the delayed input even after the delay (e.g., noise in hospital rooms \cite{park:2014:noise,siebein:2025:soundscape}, and the slowly-varying cyclic backbone of snoring, which is a non-stationary pseudo-periodic signal \cite{wang:2017:methods}). 
  \item Stochastic components: lose correlation with the delayed input due to their random nature (e.g., hypopnea, apnea \cite{kowallik:2001:breath}, and the irregular, non-stationary fluctuations of snoring that deviate from its pseudo-periodic structure \cite{hongzhang:2013:time,mikami:2015:detecting}).
\end{itemize}

Harmonic noise consists of frequency components that are integer multiples of a fundamental frequency, such as those found in medical device alarms, ventilation fan sounds, or engine noise.
ALE is particularly effective for suppressing such noise due to the strong periodicity of harmonic components, which remain highly correlated even after delay — a condition well suited to the ALE principle \cite{atkins:2021:adaptive}.

Hospital environments represent a representative use case for ALE \cite{atkins:2021:adaptive,haykin:2002:adaptive}. 
In these settings, non-stationary harmonic noises from equipment such as heart rate monitors, oxygen saturation alarms, and medical device alerts frequently interfere with speech signals used for telemedicine, staff communication, and voice-based medical record systems.

We consider the following time–frequency domain signal model:

\begin{equation}
Y(k, m) = X(k, m) + V(k, m)
\end{equation}

where $k (k = 0, 1, ..., K-1)$ is the frequency index, and $m (m = 0, 1, ..., M-1)$ is the frame index.
$Y(k, m), X(k, m)$, and $V(k, m)$ denote the STFTs of $y(n)$, $x(n)$, and $v(n)$, respectively.

\begin{equation}
\mathbf{h}(k, m) = [H_0(k, m), H_1(k, m), \ldots, H_{L-1}(k, m)]^{\mathrm{T}}
\end{equation}

\begin{equation}
Z(k, m) = \textbf{h}^{H}(k, m) \, \textbf{y}(k, m - \tau)
\end{equation}

where $Z(k, m)$ represents the output obtained by filtering the delayed input signal.
$\tau$ denotes the de-correlation delay, $\textbf{h}(k, m)$ is the adaptive filter coefficient vector, $\textbf{y}(k, m-\tau)$ is the delayed input signal vector, superscripts $^{H}$ and $^{T}$ represent the conjugate-transpose and transpose operators, respectively.
Consequently, the error signal is defined as follows:

\begin{equation}
\begin{aligned}
E(k,m) &= Y(k,m) - Z(k,m) \\
       &= Y(k,m) - \mathbf{h}^H(k,m)\mathbf{y}(k,m-\tau)
\end{aligned}
\end{equation}

where $E(\cdot)$ denotes the mathematical expectation.
The NLMS algorithm updates the filter coefficients as follows:

\begin{equation}
\label{eq:h_update}
\begin{aligned}
\mathbf{h}(k,m+1) &= \mathbf{h}(k,m) \\
                  &+ \mu(k)\frac{\mathbf{y}(k,m-\tau)\,E^{*}(k,m)}{\mathbf{y}^{H}(k,m-\tau)\mathbf{y}(k,m-\tau)+\delta} \\
\end{aligned}
\end{equation}

where $\mu(k)$ is the step size controlling the convergence speed and stability of the NLMS algorithm, $\delta$ is a small positive constant to prevent division by zero, and $E^{*}(\cdot)$ denotes the complex conjugate of $E(\cdot)$.

In traditional ALE systems, the learning rate $\mu$ is typically set as a fixed hyperparameter that controls the adaptation speed of the filter coefficients, requiring careful manual tuning to balance convergence speed and steady-state error.
However, by making $\mu$ a trainable parameter within a deep learning framework, we can enable the network to automatically learn the optimal adaptation rate for each specific signal condition and noise environment.
This approach transforms the classical ALE into a differentiable module that can be integrated into end-to-end neural architectures, where the learning rate parameter is updated through backpropagation alongside other network weights.
By treating $\mu$ as a learnable parameter, the model can dynamically adjust its adaptation behavior based on the input signal characteristics, potentially learning different learning rates for different frequency bands or time segments, which is particularly beneficial for non-stationary signals like speech or ECG (Electrocardiogram).

\begin{figure*}[htb!]
  \centering
  \includegraphics[width=0.8\linewidth]{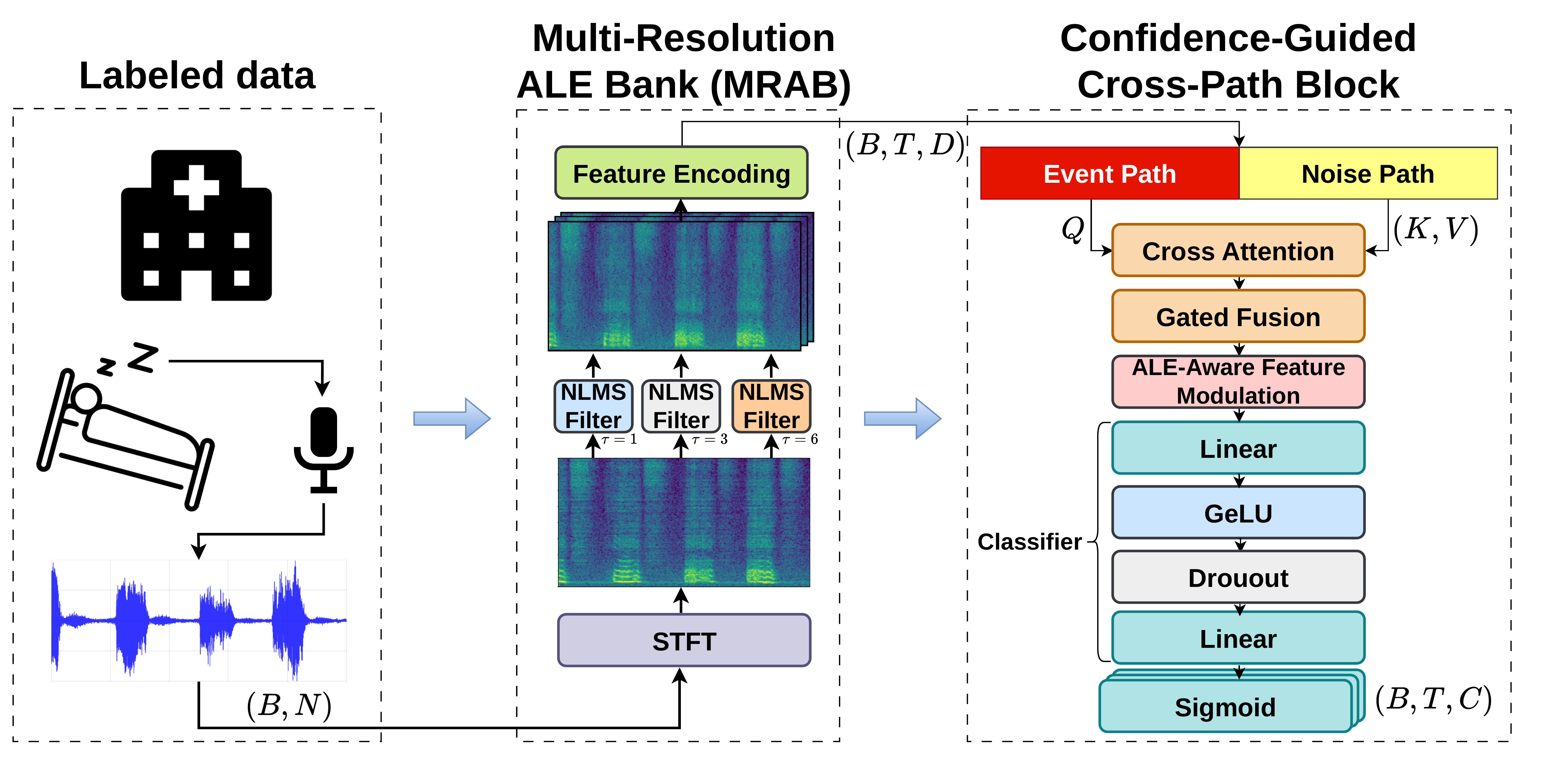}
  \caption{Architecture of the ALE-Confidence fusion framework.}
  \label{fig:fw}
\end{figure*}

\subsection{ALE-Confidence Fusion Framework}
\label{sec:acf_sed}

Having established the trainable ALE module in the preceding section, we now describe the core novelty of this work: the \textbf{ALE-Confidence Fusion Sound Event Detection} (ACF-SED) framework.
Existing ALE-based SED pipelines treat the ALE as a static pre-processing stage only the enhanced mel spectrogram $\mathbf{E}$ is forwarded to the neural network, while the filter's internal state---the per-frame confidence map $C[k,m]$---is discarded.
ACF-SED recovers this information and injects it directly into the attention and feature-modulation mechanisms of a convolutional and Transformer-based backbone, providing a principled inductive bias without additional labelled data.

The full ACF-SED pipeline is illustrated in Figure~\ref{fig:fw}.
Given a raw waveform $\mathbf{x} \in \mathbb{R}^{B \times N}$, where $B$ is the batch size and $N$ is the number of waveform samples, the system produces frame-level event probabilities $\hat{\mathbf{y}} \in[0,1]^{B \times T \times C}$,
where $T = N / h$ is the number of STFT frames with hop length $h$, which is preserved through the CNN since max-pooling is applied only along the frequency axis, and $C = 3$ is the number of target event classes (snore, hypopnea, and obstructive sleep apnea).

\begin{enumerate}
  \item \textbf{Multi-Resolution ALE Bank (MRAB):}
  three parallel NLMS filters on a shared STFT decompose the signal into event components $\{E_i\}$, noise components $\{Z_i\}$, and confidence maps $\{C_i\}$ at short-, medium-, and long-term scales.
  \item \textbf{Dual-Stream CNN:}
  independent convolutional encoders map the stacked enhanced mel $\mathbf{E}_{\mathrm{mel}}$ and noise mel $\mathbf{Z}_{\mathrm{mel}}$ into $D$-dimensional frame sequences $\mathbf{e}_{\mathrm{seq}}$ and $\mathbf{n}_{\mathrm{seq}}$.
  \item \textbf{Learnable Confidence Pooler:}
            a frequency-and-resolution attention pooler collapses the $(3\!\times\!F)$-dimensional confidence tensor into a scalar confidence trace $\mathbf{c} \in \mathbb{R}^{B \times T}$.
  \item \textbf{Confidence-Guided Cross-Path Block (CCP):}
  stacked blocks process the event and noise streams through two asymmetric parallel paths.
  The \emph{event path} applies Symmetric Confidence-Biased Multi-Head Attention (Sym-CB-MHA), which injects the ALE confidence trace $c$ into the attention score matrix, followed by a convolutional sub-layer and ALE-Aware Feature Modulation (AFM).
  The \emph{noise path} applies a lightweight feed-forward sub-layer, providing contextual cues without modelling events independently.
  The two paths are then fused via cross-attention with a learned frame-wise gate, allowing the event stream to selectively incorporate noise context at ambiguous event boundaries.
  \item \textbf{Frame Classifier:}
  a two-layer MLP with sigmoid activation produces per-class predictions.
\end{enumerate}

\subsubsection{Multi-Resolution ALE Bank (MRAB)}
\label{sec:mrab}

The decorrelation delay $\tau$ in the NLMS filter determines which periodic components are suppressed: a filter with delay $\tau$ removes signal components whose autocorrelation remains high at lag $\tau \cdot t_{\mathrm{hop}}$, where $t_{\mathrm{hop}} = 10$\,ms is the STFT frame shift.
Different noise sources in sleep audio exhibit distinct periodicity time scales:

\begin{itemize}
  \item \textbf{Short-term correlations ($\tau{=}1$):}
  Rapid oscillatory components such as high-frequency harmonic noise from medical equipment (e.g.\ ventilators, cardiac monitors) and the upper harmonics of snoring waveforms maintain strong correlation at very short delays.
  A filter with $\tau{=}1$ is sensitive to these fast-varying periodic structures and effectively isolates transient snoring onsets from the background.
  
  \item \textbf{Medium-term correlations ($\tau{=}3$):}
  The fundamental frequency of snoring typically falls in the range 60-300\,Hz.
  At $\tau{=}3$ frames, the filter aligns with the periodicity of mid-frequency snoring components and captures background breathing harmonics that persist at this time scale, which a short-delay filter would miss.
  
  \item \textbf{Long-term correlations ($\tau{=}6$):}
  Low-frequency noise components arising from slow respiratory rhythm and airway turbulence maintain coherence over longer intervals.
  A study \cite{atkins:2021:adaptive} demonstrated that non-stationary harmonic noise in clinical environments contains slowly varying periodicities that require longer decorrelation delays to suppress effectively.
  The $\tau{=}6$ filter targets these components, providing a clean residual that more faithfully represents aperiodic apnea events.
  \end{itemize}

No single $\tau$ value can simultaneously suppress noise across all three time scales.
Using only $\tau{=}1$ leaves medium- and low-frequency periodic noise in the residual, contaminating the confidence map with false high-confidence values.
Using only $\tau{=}6$ fails to capture fast-oscillating snoring components, causing them to leak into the aperiodic residual and inflating false apnea detections.
MRAB addresses this by running all three filters in parallel on the shared STFT and aggregating their confidence maps through the Learnable Confidence Pooler, enabling the model to exploit complementary noise-removal coverage across temporal scales.
A single NLMS filter with fixed decorrelation delay $\tau$ is sensitive only to periodicities at the corresponding time scale.
Sleep-disordered breathing events span a wide range of durations, so a single filter is unlikely to be an ideal match for all of them.
MRAB addresses this by running three filters in parallel on the same STFT coefficients $Y[k,m]$:

\begin{equation}
  \{E_i,\,Z_i\} = \mathrm{NLMS}_i(Y), \quad i \in \{1,2,3\},
\end{equation}

with parameters $(\tau_1,L_1)=(1,8)$, $(\tau_2,L_2)=(3,15)$, and $(\tau_3,L_3)=(6,24)$, covering short-, medium-, and long-term correlations respectively.
Each filter's step size $\mu_i = \sigma(\mu_{i,\mathrm{logit}})$ is independently learnable via back-propagation through the cross-entropy loss.
The STFT is computed once and shared across all filters, keeping the computational overhead negligible.
The per-filter confidence map is defined as

\begin{equation}
  C_i[k,m] = \frac{|E_i[k,m]|^2}{|E_i[k,m]|^2 + |Z_i[k,m]|^2 + \epsilon}
  \;\in [0,1],
  \label{eq:confidence}
\end{equation}

where $C_i \to 1$ indicates a predominantly aperiodic (event-like) frame and $C_i \to 0$ a periodic (noise-like) frame.
Mel spectrograms of $E_i$ and $Z_i$ are stacked along a resolution channel, yielding $\mathbf{E}_{\mathrm{mel}},\mathbf{Z}_{\mathrm{mel}} \in \mathbb{R}^{B \times 3 \times F \times T}$.

\subsubsection{Dual-Stream CNN Feature Extractor}
\label{sec:dual_cnn}

Two independent three-stage convolutional encoders---the \emph{Event CNN} and the lighter \emph{Noise CNN}---process $\mathbf{E}_{\mathrm{mel}}$ and $\mathbf{Z}_{\mathrm{mel}}$ respectively.
Each stage consists of \texttt{Conv2d}--\texttt{BatchNorm}--\texttt{GELU}--\texttt{MaxPool}(2$\times$1) layers that progressively reduce the frequency dimension while preserving the time axis.
An adaptive average pooling layer collapses the remaining frequency dimension to one, producing $\mathbf{e}_{\mathrm{seq}}, \mathbf{n}_{\mathrm{seq}} \in \mathbb{R}^{B \times T \times D}$ with $D = 128$.

\subsubsection{Learnable Confidence Pooler}

 The three per-resolution confidence maps $C_r \in \mathbb{R}^{B \times F \times T}$ ($r \in \{1,2,3\}$) are collapsed into a single temporal confidence trace
  $c \in \mathbb{R}^{B \times T}$ by taking a weighted sum over both the                                                                                       
  resolution and frequency axes.
  The resolution weight $w_{\mathrm{res}} \in \mathbb{R}$ controls how
  much each ALE scale contributes, and the frequency weight
  $w_{\mathrm{freq}} \in \mathbb{R}$ controls the contribution of each
  mel bin; both are obtained by applying softmax to learnable logit vectors
  initialised to zero and jointly optimised with the
  classification objective:
  \begin{equation}
    c[b,\,t]
    = \sum_{r=1}^{3} w_{\mathrm{res}}
      \sum_{f=1}^{F} w_{\mathrm{freq}} \cdot C_{r}[b,\,f,\,t],
    \label{eq:pooler}
  \end{equation}
  where $b$, $f$, and $t$ index the batch sample, mel frequency bin, and
  time frame, respectively.
  This allows the model to up-weight the ALE resolution and frequency bands
  most discriminative for each event class, which a simple average cannot achieve.

\subsubsection{Confidence-Guided Cross-Path Block (CCP)}                                                                                                     
\label{sec:ccp}                                                                                                                             

The Confidence-Guided Cross-Path Block (CCP) forms the core sequence modelling unit of ACF-SED.
Unlike conventional Conformer-based SED systems that process a single enhanced stream, each CCP block maintains two parallel paths an \emph{Event Path} and a \emph{Noise Path} and fuses them through gated cross-attention.
The underlying motivation is that distinguishing an apnea from a silent breathing pause is substantially easier when the model can simultaneously observe the concurrent snoring or background-breathing pattern encoded by the noise path.\\

\par\noindent\textbf{{Symmetric Confidence-Biased Multi-Head Attention (Sym-CB-MHA)}}\par

Sym-CB-MHA injects the ALE confidence trace $\mathbf{c} \in \mathbb{R}^{B \times T}$ into the attention score matrix by first projecting it to a per-head bias:

\begin{equation}
  \mathrm{cb}[b,\,h,\,t] = W_c\,c[b,\,t],
\label{eq:cb_proj}
\end{equation}
  
where $b$, $h$, and $t$ index the batch sample, attention head, and time frame, respectively.
The bias is then added symmetrically to the standard dot-product score:

\begin{equation}
    \mathrm{scores}[b,\,h,\,q,\,k]
    = \frac{\mathbf{Q}_{q}\,\mathbf{K}_{k}^{\top}}{\sqrt{d_k}}
    + \alpha\bigl(\mathrm{cb}[b,\,h,\,q] + \mathrm{cb}[b,\,h,\,k]\bigr),
\label{eq:sym_cb_mha}
\end{equation}
  
where $q$ and $k$ are the query and key time-frame indices, $\mathbf{Q}_{q},\,\mathbf{K}_{k} \in \mathbb{R}^{d_k}$ are the projected query and key vectors at those positions, $d_k$ is the per-head key dimension, and $\alpha = \sigma(\alpha_{\mathrm{logit}})$ is a learnable scalar gate.
The symmetric bias $\mathrm{cb}[b,h,q]+\mathrm{cb}[b,h,k]$ simultaneously raises scores when the query position, the key position, or both are event-likely, so that co-occurring high-confidence frames—such as those within a sustained apnea—receive a double boost that reinforces long-range intra-event dependencies.\\

\par\noindent\textbf{Event Path.}\par
The event feature sequence $\mathbf{e} \in \mathbb{R}^{B \times T \times D}$ is processed through the sandwich-style block \cite{gulati:2020:conformer}, with the self-attention module replaced by Sym-CB-MHA:

\begin{align}
      \mathbf{e_0} &\leftarrow \mathbf{e_{in}}
          + \tfrac{1}{2}\,\mathrm{FFN}_{1}(\mathbf{e_{in}}), \nonumber \\
      \mathbf{e_1} &\leftarrow \mathbf{e_0}
          + \mathrm{Sym\mbox{-}CB\mbox{-}MHA}\bigl(\mathrm{LN}(\mathbf{e_0}),\,\mathbf{c}\bigr),
      \label{eq:ccp_mha} \\
      \mathbf{e_2} &\leftarrow \mathbf{e_1}
          + \mathrm{ConvModule}(\mathbf{e_1}), \nonumber \\
      \mathbf{e_3} &\leftarrow \mathbf{e_2}
          + \tfrac{1}{2}\,\mathrm{FFN}_{2}(\mathbf{e_2}), \nonumber\\
            \mathbf{e_{out}} &\leftarrow \mathrm{LN}\bigl(\mathbf{e_3}\bigr), \nonumber    
\end{align}

where each residual term is subject to DropPath during training.
The feed-forward modules $\mathrm{FFN}_{1,2}$ follow a pre-norm design with expansion factor~4 and SiLU activation.
$\mathrm{ConvModule}$ consists of layer normalisation, a pointwise expansion (Linear $D \to 2D$) followed by GLU, a depthwise convolution of kernel size 31, batch normalisation, SiLU activation, and a pointwise projection back to $D$ dimensions.\\

\par\noindent\textbf{Noise Path.}\par

The noise feature sequence $\mathbf{n} \in \mathbb{R}^{B \times T \times D}$ is processed by a single lightweight feed-forward sub-layer with expansion factor:

\begin{equation}
  \mathbf{n_{out}} \leftarrow \mathrm{LN}\bigl(
          \mathbf{n_{in}} + \mathrm{FFN}_{\mathrm{noise}}(\mathbf{n_{in}})\bigr).
\label{eq:noise_path}
\end{equation}

The deliberately reduced capacity reflects the auxiliary role of the noise path: it provides contextual cues to the event path rather than modelling events independently, and allocating fewer parameters avoids overfitting on the less discriminative noise signal.\\

\par\noindent\textbf{Cross-Attention Gated Fusion.}\par
 After the two paths are independently updated, the event stream queries the                                                                                  
  noise stream via scaled dot-product cross-attention with $\textbf{H/2}$                                                                               
  heads—half the number used in self-attention—to reduce cross-path parameter
  cost:
  
\begin{equation}
\begin{aligned}
\mathbf{e}_{\mathrm{cross}} &= \mathrm{CrossAttn}(Q,\,K,\,V), \\
Q &= \mathrm{LN}(\mathbf{e_{out}}), \\
K &= V = \mathrm{LN}(\mathbf{n_{out}}).
\end{aligned}
\label{eq:cross_attn}
\end{equation}
  
  A learned frame-wise gate then interpolates between the event
  self-representation and the retrieved noise context:
  
\begin{align}
  \mathbf{g} &= \sigma\bigl(
      W_{g}\,[\mathbf{e_{out}}\,;\,\mathbf{e}_{\mathrm{cross}}]\bigr),
  \\
  \mathbf{e_{fus}} &\leftarrow \mathbf{e_{out}} \odot \mathbf{g}
    + \mathbf{e}_{\mathrm{cross}} \odot (1 - \mathbf{g}),
\label{eq:gated_fusion}
\end{align}
  
  where $[\,\cdot\,;\,\cdot\,]$ denotes concatenation along the feature axis,
  $W_g \in \mathbb{R}^{D \times 2D}$, $\odot$ is the Hadamard product,
  and $\sigma(\cdot)$ is the sigmoid function.
  The gate $g \in (0,1)^{B \times T \times D}$ produces a per-frame,
  per-channel mixing coefficient: it approaches one on clearly identifiable
  event frames and approaches zero at ambiguous boundaries where noise context
  is informative.\\

\par\noindent\textbf{ALE-Aware Feature Modulation (AFM).}\par
Following gated fusion, AFM amplifies the fused event features according to the ALE confidence:

\begin{equation}
      \mathbf{e_{afm}} \leftarrow \mathbf{e_{fus}}
          \odot \bigl(\mathbf{1} + \beta \cdot \sigma(W_{\mathrm{mod}}\,\mathbf{c})\bigr),
\label{eq:afm}
\end{equation}

where $W_{\mathrm{mod}} \in \mathbb{R}^{D \times 1}$ projects the scalar confidence to a $D$-dimensional channel-wise scale, and $\tau = \sigma(\tau_{\mathrm{logit}})$ is a learnable temperature.
The offset ensures the scale is at least unity:
high-confidence (event-likely) frames are \emph{amplified}, while low-confidence (noise-like) frames are passed through \emph{unchanged}.

\subsubsection{Frame Classifier}
\label{sec:classifier}

After CCP blocks, the event feature sequence $\mathbf{e} \in \mathbb{R}^{B \times T \times D}$ is passed through a two-layer MLP with an intermediate dropout:

\begin{equation}
      \hat{\mathbf{y}} = \sigma\bigl(
          W_{2}\,\mathrm{GELU}(W_{1}\,\mathbf{e_{afm}} + b_{1}) + b_{2}
      \bigr),
      \label{eq:classifier}
\end{equation}

where $W_1 \in \mathbb{R}^{D/2 \times D}$ and $W_2 \in \mathbb{R}^{C \times D/2}$ are learned weights, $C{=}3$ denotes the number of event classes (snoring, hypopnea, obstructive apnea), and $\hat{\mathbf{y}} \in [0,1]^{B \times T \times C}$ gives independent per-class frame-level probabilities.

\section{Experimental Setup}
We train the model on the Intel Xeon-Silver 4410Y CPU and NVIDIA GeForce RTX 5090 GPU.
All signals were sampled at 4,000 Hz before processing, and features are extracted using a 25 ms analysis window with a 10 ms hop length.
The model is trained with AdamW \cite{loshchilov:2017:decoupled} using an initial learning rate of 1e-4 and a OneCycleLR scheduler \cite{smith:2019:super}; the initial scale is set to 0.001.
To stabilize training, gradient clipping is applied with a norm of 1.0, and the model is trained for 100 epochs with a batch size of 32.
A summary of these hyperparameters is provided in Table~\ref{tab:exp_conditions}.

\begin{table}[htb!]
\centering
\begin{tabular}{ll}
\hline
\hline
Sampling rate               & 4,000~Hz \\
Window length               & 25~ms \\
Hop length                  & 10~ms \\
Learning rate               & 1$e$-4 \\
Initial scale               & 0.001 \\
Gradient clipping norm      & 1.0 \\
Epoch                       & 100 \\
Batch size                  & 32 \\
Optimizer          & AdamW \\
Scheduler          & OneCycleLR \\
\hline
\hline
\end{tabular}
\caption{Experimental Conditions.}
\label{tab:exp_conditions}
\end{table}

\subsection{Dataset}
The dataset used in our experiment, the Audio-Polysomnography Sleep Apnea Analysis (APSAA) \cite{gonzalez:2024:audio}, provides synchronized full-night audio and polysomnography recordings from 32 participants, along with manually annotated events labeled from the polysomnography studies.
All participants were informed in detail about the research, and written consent was obtained from those who agreed to participate.
Recordings were collected in the sleep laboratory of Doctor Sagaz Hospital in Jaén, Spain, between September 2021 and April 2022.

This dataset was labeled using the Embla RemLogic 3.1 software\footnote{\url{https://neurolite.ch/en/products/sleep-medicine/embla\%C2\%AE-remlogic\%E2\%84\%A2-software}}\cite{gonzalez:2025:polygraph}.
To ensure compliance with the American Academy of Sleep Medicine (AASM) guidelines \cite{chang:2023:international}, a manual labeling step was conducted under medical supervision based on the automatic detection of respiratory events generated by RemLogic.
The audio recorder (ZOOM H4n Pro) was positioned approximately 30~cm above the participant’s head to capture sound signals sampled at 44,100~Hz.
To reduce data size, the signals were converted to a single channel and downsampled to 4,000~Hz.

For model training and evaluation, the dataset was split into train, validation, and test sets at a ratio of 8:1:1, consisting of 25, 3, and 4 subjects, respectively.

\begin{figure}[h!]
  \centering
  \includegraphics[width=\linewidth]{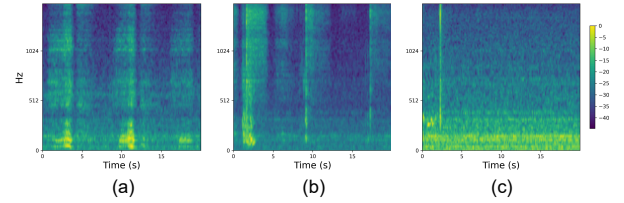}
  \caption{Visualization of spectrograms for three key sleep audio events.
  (a) Snoring, (b) Hypopnea, (c) Obstructive sleep apnea (OSA).}
  \label{fig:sha}
\end{figure}

In this study, the detection targets in sleep audio are snoring, hypopnea, and OSA.
Their definitions are as follows, and Figure~\ref{fig:sha} shows the spectrograms of each event.

\begin{itemize}
    \item \textbf{Snoring} is an abnormal respiratory sound that occurs during sleep, caused by vibrations of upper airway tissues (such as the uvula, soft palate, and tongue base) and partially obstructed airflow.
    It predominantly occurs during inspiration.
    The snoring ratio was defined as the cumulative snoring time divided by the total sleep time.
    A snoring episode was defined as three or more consecutive snoring events, and the snoring episode index was defined as the number of snoring episodes per hour \cite{kim:2022:definition}.
    \item \textbf{Hypopnea} refers to abnormally shallow breathing or an unusually low respiratory rate \cite{gould:1988:sleep}.
    It is generally defined as a reduction in the volume of air entering the lungs and can lead to hypoxemia (decreased blood oxygen levels).
    Hypopnea during sleep is classified as a sleep disorder. Moderate to severe hypopnea disrupts sleep, preventing patients from feeling refreshed even after obtaining sufficient sleep duration.
    The breathing disturbance causes decreased blood oxygen levels, which can disrupt sleep stages.
    \item \textbf{Obstructive sleep apnea} (OSA) is a sleep-related breathing disorder where airflow decreases or completely stops despite breathing efforts during sleep \cite{mcnicholas:2008:diagnosis}.
    This occurs when relaxed muscles cause soft tissue collapse in the throat, blocking the upper airway.
    Episodes of partial (hypopnea) or complete (apnea) breathing cessation last at least 10 seconds, typically 10-30 seconds, though some exceed one minute.
    These events can rapidly decrease blood oxygen saturation, sometimes dropping by more than 40\% in severe cases.
\end{itemize}

\subsection{Evaluation Metrics}
In this study, we adopt multiple evaluation metrics to comprehensively assess the performance of sound event detection (SED).
Specifically, we employ the event-based F1-score, segment-based F1-score, Error Rate (ER), and Polyphonic Sound Detection Score (PSDS) to capture different aspects of detection accuracy and temporal precision.

The proposed PSDS \cite{bilen:2020:framework} is based on the intersection between detected sound events and ground truth without strict event-based requirements.
A detection is considered a true positive (TP) if it meets the following two criteria:
The Detection Tolerance Criterion (DTC) does not consider a detection as a TP if the intersection between the detected event and the ground truth is smaller than $\rho_{DTC}$ relative to the detected event length.
The Ground Truth Criterion (GTC) measures the intersection between the ground truth event and the set of detections that satisfy the DTC.
A detection is regarded as a TP if this intersection, normalized by the ground truth event length, is equal to or greater than $\rho_{GTC}$.
The Cross-Trigger Tolerance Criterion (CTTC) is defined to account for cross-triggers given a threshold $\rho_{cttc}$.
The penalty cost for user experience is controlled by a weighting parameter $\alpha_{CT}$.
Although cross-triggers are generally more identifiable than false positives (FP), they may still lead to a more negative user experience.
The overall Polyphonic Sound Detection Receiver Operating Characteristic (PSD-ROC) curve is computed by averaging the class-wise ROC curves while considering the class instability cost ($\alpha_{ST}$).
Finally, the PSDS score is obtained by integrating the PSD-ROC curve from 0 to a specific parameter $e_{max}$.
The parameter $e_{max}$, measured in events per hour, defines the maximum false positive rate at which the operating point remains relevant.

\section{Results and Discussion}
\subsection{Comparison with Baseline Models}

Table~\ref{tab:main_results} reports the Event-F1, Segment-F1, and PSDS scores of ACF-SED against four representative baselines on the APSAA evaluation set.
ACF-SED achieves an Event-F1 of 0.7105, Segment-F1 of 0.6930, and PSDS of 0.7633, outperforming all baselines across every metric.

\begin{table}[htb!]
\centering
\begin{tabular}{cccc}
\hline
\textbf{Method} &  \textbf{Event-F1} & \textbf{Segment-F1} & \textbf{PSDS}  \\
\hline
CRNN & 0.6148 & 0.6169 & 0.7175  \\
Conformer & 0.5975 & 0.6109 & 0.5015  \\
BEATs + Conformer & 0.6364 & 0.6071 & 0.6949  \\
AST & 0.6710 & 0.6746 & 0.7435  \\
\hline
Ours & \textbf{0.7105} & \textbf{0.6930} & \textbf{0.7633}  \\
\hline
\end{tabular}
\caption{Performance comparison on the APSAA evaluation set.}
\label{tab:main_results}
\end{table}

The plain Conformer baseline yields the lowest PSDS (0.5015), indicating that transformer-based sequence modelling alone is insufficient without an appropriate signal-processing front-end for this domain.
CRNN achieves a competitive PSDS of 0.7175, benefiting from its joint convolutional recurrent design, yet falls short on Event-F1 (0.6148) due to its limited temporal precision.
Augmenting the Conformer with BEATs pre-training raises Event-F1 to 0.6364 but does not match CRNN in PSDS, suggesting that generic audio pre-training provides limited benefit for the narrow domain of sleep audio.
AST achieves the strongest baseline performance (Event-F1 0.6710, PSDS 0.7435) through its pure attention mechanism.

\subsection{AHI Estimation}
To assess the clinical utility of ACF-SED, we derive the Apnea--Hypopnea Index (AHI) solely from the frame-level predictions of the three sound event classes—snoring, hypopnea, and obstructive apnea—produced by the audio classifier, without any physiological sensor input.
Detected apnea and hypopnea events are counted from the predicted frame labels and divided by the total recording duration in hours to yield the estimated AHI.
Table~\ref{tab:ahi_results} compares the estimated AHI against the reference AHI obtained from polygraphy for four representative test subjects.

\begin{table}[htbp]
\centering
\begin{tabular}{lcccc}
\toprule
\multirow{2}{*}{Subject} & \multicolumn{2}{c}{AHI} & \multicolumn{2}{c}{Severity} \\
\cmidrule(lr){2-3} \cmidrule(lr){4-5}
 & Pred & GT & Pred & GT \\
\midrule
20211019A & 19.05 & 40.79 & Moderate & Severe \\
20211107B & 31.59 & 39.11 & Severe & Severe \\
20220116B & 39.61 & 66.31 & Severe & Severe \\
20220130C & 18.40 & 27.82 & Moderate & Moderate \\
\bottomrule
\end{tabular}
\caption{AHI estimation and OSA severity classification results on the APSAA evaluation set.}
\label{tab:ahi_results}
\end{table}

\begin{figure}[htb!]
  \centering
  \subfloat[]{
    \includegraphics[width=1\linewidth]{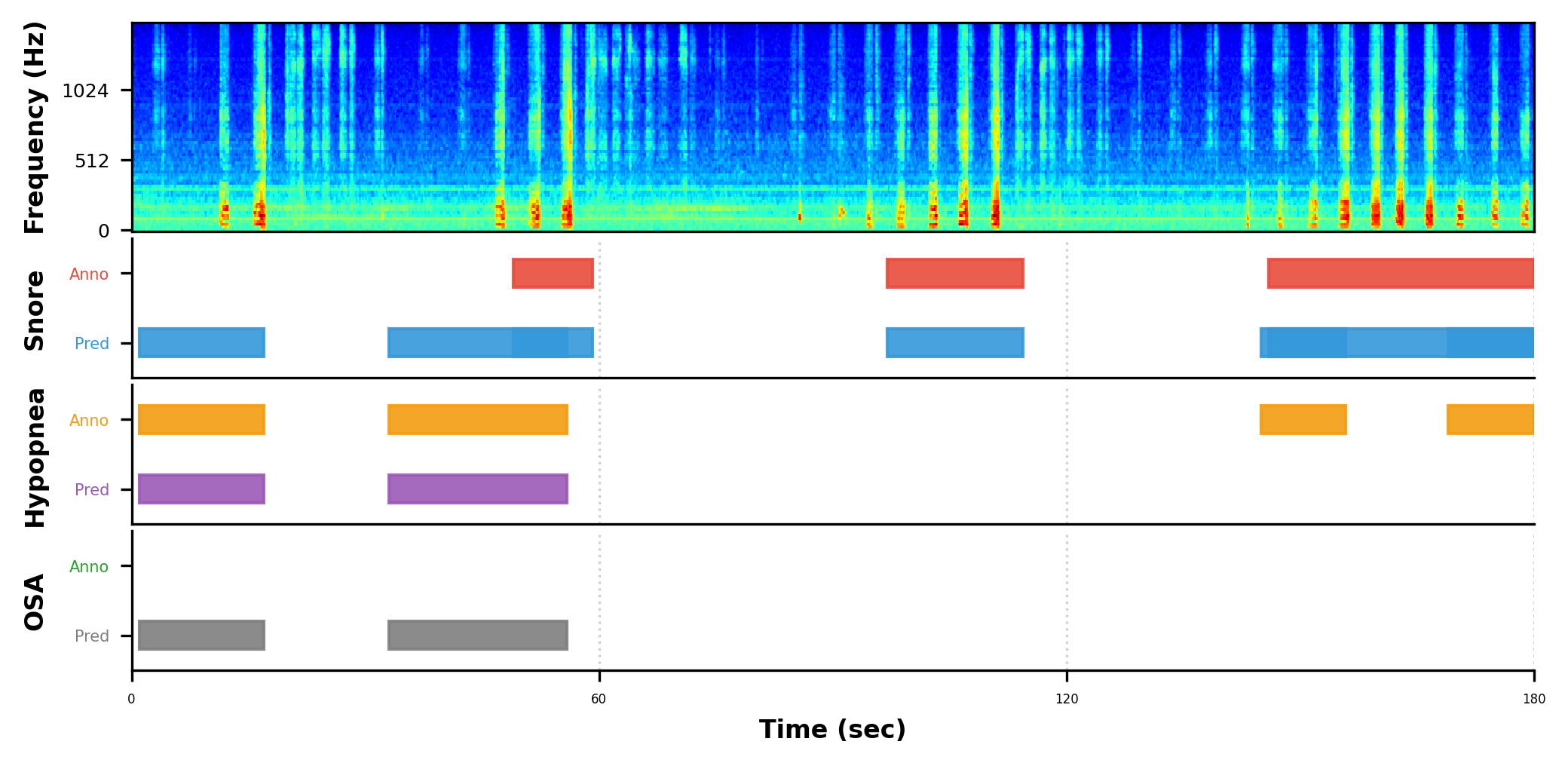}
    \label{fig:cnn_a}
  }
  \hfil
  \subfloat[]{
    \includegraphics[width=1\linewidth]{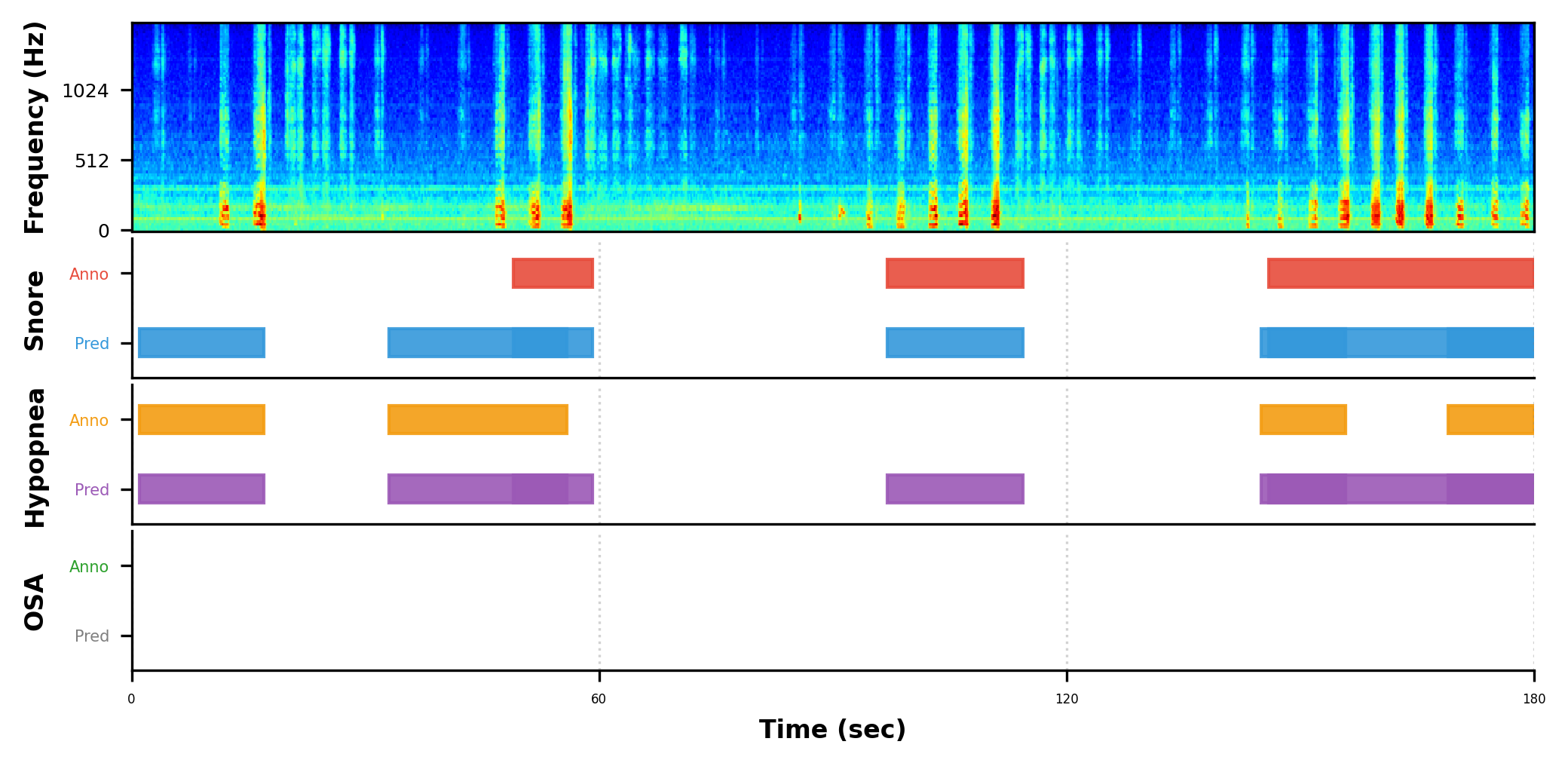}
    \label{fig:cnn_b}
  }
  
  \caption{Qualitative comparison of frame-level sleep sound event
               detection on a representative test recording.
               (a) predictions from AST \cite{gong:2021:ast}.
               (b) predictions from the proposed ACF-SED.
  }
  \label{fig:detection_comparison}
\end{figure}

Figure~\ref{fig:detection_comparison} presents a qualitative comparison of frame-level event detection results between the AST and the proposed ACF-SED on a representative test recording.
The upper panel shows the predictions produced by AST, and the lower panel shows those of ACF-SED, with the ground-truth event boundaries indicated by the shaded regions.
As shown in the Figure~\ref{fig:cnn_a}, AST tends to produce fragmented predictions within event intervals and generates spurious activations during silent or background-breathing segments, reflecting its lack of a signal-processing prior to distinguish periodic from aperiodic audio components.
In contrast, the lower panel demonstrates that ACF-SED yields more contiguous and temporally precise detections: event boundaries align more closely with the ground truth, false activations during non-event frames are substantially reduced, and the three event classes---snoring, hypopnea, and obstructive apnea---are more cleanly separated.
These qualitative observations are consistent with the quantitative improvements reported in Table~\ref{tab:main_results}.

\subsection{Ablation Study}
\par\noindent\textbf{Effect of Learnable Step Size}\par

The gap is particularly pronounced on Event-F1, which is sensitive to temporal precision: a fixed step size calibrated for the average signal condition is suboptimal for the diverse durations of sleep events (0.5--60\,s), whereas a learnable $\mu$ adapts the filter convergence rate to the distribution of events observed during training.
These results validate the design choice of making the NLMS step size differentiable and jointly optimised with the downstream classification objective.

\begin{figure}[htb!]
  \centering
  \includegraphics[width=1\linewidth]{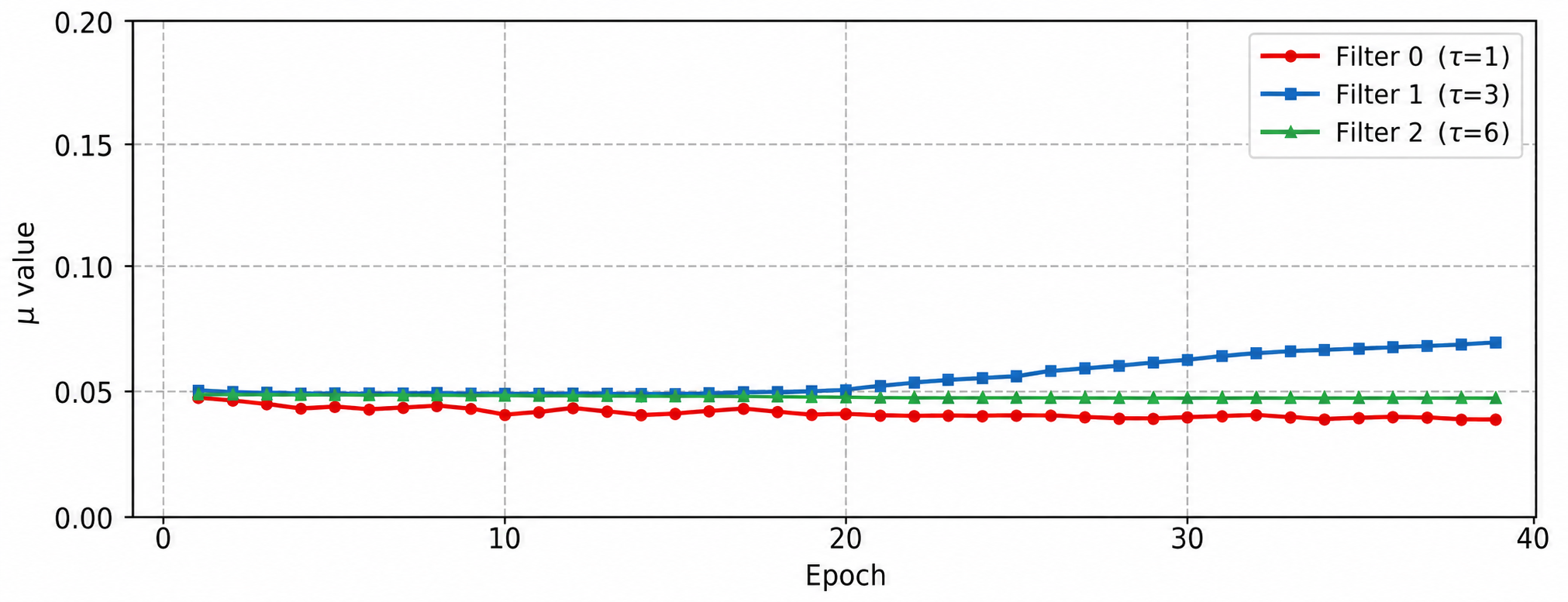}
  \caption{Convergence of the three learnable NLMS step sizes $\mu_1$, $\mu_2$, $\mu_3$ during training.
  Each $\mu_i$ is initialised to $0.05$ and jointly optimised with the classification objective.}
  \label{fig:mu_training}
\end{figure}

As shown in Figure~\ref{fig:mu_training}, the three step sizes converge to distinct values, reflecting the differing adaptation requirements of each temporal scale.
The short-term filter ($\mu_1$, $\tau{=}1$) converges to a relatively large step size, enabling rapid adaptation to transient snoring events.
The medium-term filter ($\mu_2$, $\tau{=}3$) settles at an intermediate value, balancing convergence speed and steady-state error for standard-duration events.
The long-term filter ($\mu_3$, $\tau{=}6$) converges to the smallest step size, prioritising stability over the extended time scales associated with apnea and hypopnea events.
This differentiation would not be achievable with a single fixed $\mu$ shared across all filters, further motivating the multi-resolution design of MRAB and the learnable parameterisation of each filter's adaptation rate.\\

\par\noindent\textbf{Effect of Delay Parameter}\par
Table~\ref{tab:tau_ablation} reports the effect of the delay parameter $\tau$
  on detection performance, comparing three uniform configurations---where all                                                                
  filters share the same $\tau \in \{1, 3, 6\}$---against the proposed
  multi-resolution assignment $\tau{=}\{1,3,6\}$.

  Among the uniform settings, $\tau{=}1$ yields the highest Event-F1 (0.7016)
  and Segment-F1 (0.6981), suggesting that a short delay facilitates rapid
  adaptation to transient events such as snoring.
  However, its PSDS (0.7487) falls short of the multi-resolution baseline,
  indicating that a single short delay is insufficient to capture the temporal
  diversity of sleep-disordered breathing events, which range from brief snoring
  bursts to prolonged apnea episodes lasting tens of seconds.
  Increasing the delay to $\tau{=}3$ or $\tau{=}6$ degrades all three metrics,
  as a uniform long delay slows convergence and impairs the detection of
  short-duration events.

  In contrast, the multi-resolution configuration $\tau{=}\{1,3,6\}$ assigns
  distinct delays to each filter, enabling simultaneous adaptation at short,
  medium, and long temporal scales.
  This heterogeneous assignment achieves the best PSDS score (0.7633),
  outperforming all uniform baselines.
  The improvement in PSDS, which directly measures temporal localisation
  accuracy, confirms that the three filters capture complementary spectral
  dynamics that no single $\tau$ value can cover alone.
  Although the Event-F1 of the multi-resolution setting matches that of
  $\tau{=}1$ (both 0.7016), the superior PSDS demonstrates that
  multi-resolution delay assignment yields more precise event boundary
  estimates, validating the design rationale of the Multi-Resolution
  ALE Bank (MRAB).\\

\begin{table}[htb!]
\centering
\begin{tabular}{cccc}
\toprule
$\tau$ & Event-F1 & Segment-F1 & PSDS \\
\midrule
1  & \textbf{0.7016} & \textbf{0.6981} & 0.7487 \\
3  & 0.6663 & 0.6618 & 0.7308 \\
6  & 0.6733 & 0.6625 & 0.7229 \\
\midrule
1,3,6 & \textbf{0.7105} & 0.6877 & \textbf{0.7633} \\
\bottomrule
\end{tabular}
\caption{Effect of the delay parameter $\tau$ in the multi-resolution ALE bank on the APSAA evaluation set.                               
             Rows with a single $\tau$ value denote configurations where all three filters share the same delay;
             $\tau{=}\{1,3,6\}$ assigns a distinct delay to each filter.}
\label{tab:tau_ablation}
\end{table}

\par\noindent\textbf{Comparison of STFT-Domain and Time-Domain ALE}\par

\begin{figure}[htb!]
  \centering
  \includegraphics[width=1\linewidth]{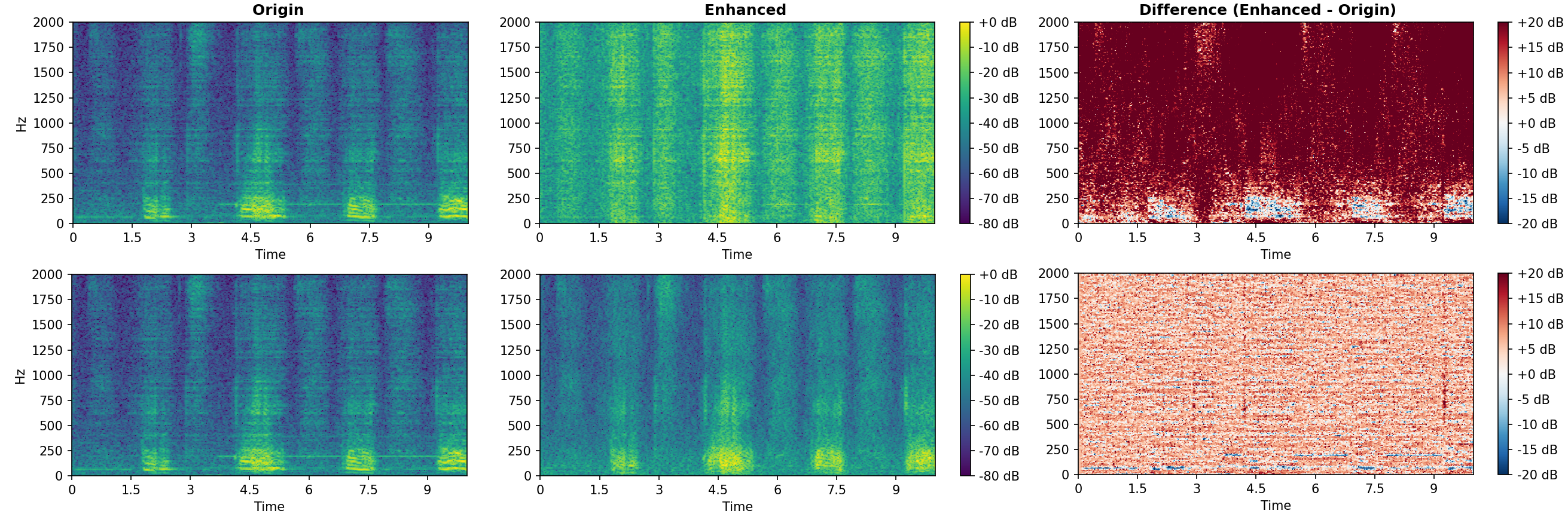}
  \caption{Comparison of spectrograms obtained from ALE processing in (top) time domain and (bottom) STFT domain, showing the original sound, enhanced output, and residual difference (from left to right).
}
  \label{fig:tvsst}
\end{figure}

In this experiment, we compare the noise removal performance of ALE applied in both the time domain and the STFT domain.
The time domain results shown in the upper panel of Figure~\ref{fig:tvsst} reveal that the difference map clearly exhibits structural patterns of the original signal, suggesting that signal loss or distortion occurred during the noise removal process.
In contrast, the difference map in the STFT domain results shown in the lower panel displays only irregular noise patterns, confirming that the STFT-based approach is more effective at selectively suppressing background noise while preserving the integrity of the original signal information.

\section{Conclusion}
 In this study, we proposed ACF (Adaptive Line Enhancer Confidence Fusion), a novel audio-based framework for detecting snoring,      
  hypopnea, and apnea events in sleep studies. By integrating an Adaptive Line Enhancer (ALE) with a convolutional and Transformer-based architecture, the
  proposed model effectively enhances periodic respiratory signals while suppressing background noise, enabling robust sleep event    
  detection from raw audio. The estimated AHI derived from detected events demonstrated clinically reliable accuracy. Experimental    
  results show that ACF outperforms existing models including CRNN, Conformer, and AST in both F1-score and PSDS, validating the      
  effectiveness of combining ALE-based signal enhancement with the the CNN’s feature extraction and the attention mechanism’s ability to model temporal dependencies.
  These findings suggest that ACF is a promising and accessible alternative to conventional polysomnography for sleep-disordered breathing assessment.

\newpage
\section{Acknowledgments}
This work was supported in part by: the Institute of Information \& Communications Technology Planning \& Evaluation (IITP) grant funded by the Korean government (MSIT) under Grant No. RS-2019-II190079 for the Artificial Intelligence Graduate School Program at Korea University; the Institute of Information \& Communications Technology Planning \& Evaluation (IITP) grant funded by the Korean government (MSIT) under Grant No. RS-2025-02304828 for the Artificial Intelligence Star Fellowship Support Program to Nurture the Best Talents;
the Institute of Information \& Communications Technology Planning \& Evaluation (IITP) grant funded by the Korean government (MSIT) under Grant No. RS-2025-25442867;
the National Research Foundation of Korea (NRF) grant funded by the Korean government (MSIT) under Grant No. RS-2025-24535409;
the Technology Development Program funded by the Ministry of SMEs and Startups (MSS, Korea) under Grant No. RS-2026-25534256;
and the Supreme Prosecutor's Office Research Grant in 2026 (research title: Development of fake voice detection technology robust in new voice generation technology and speaker recognition).
Acknowledgments and references may begin on an earlier page if space permits.

\section{Generative AI Use Disclosure}
Generative AI was used to refine the manuscript’s wording and correct grammatical errors.

\bibliographystyle{IEEEtran}
\bibliography{mybib}

\begin{thebibliography}{10}
\providecommand{\url}[1]{#1}
\csname url@samestyle\endcsname
\providecommand{\newblock}{\relax}
\providecommand{\bibinfo}[2]{#2}
\providecommand{\BIBentrySTDinterwordspacing}{\spaceskip=0pt\relax}
\providecommand{\BIBentryALTinterwordstretchfactor}{4}
\providecommand{\BIBentryALTinterwordspacing}{\spaceskip=\fontdimen2\font plus
\BIBentryALTinterwordstretchfactor\fontdimen3\font minus
  \fontdimen4\font\relax}
\providecommand{\BIBforeignlanguage}[2]{{%
\expandafter\ifx\csname l@#1\endcsname\relax
\typeout{** WARNING: IEEEtran.bst: No hyphenation pattern has been}%
\typeout{** loaded for the language `#1'. Using the pattern for}%
\typeout{** the default language instead.}%
\else
\language=\csname l@#1\endcsname
\fi
#2}}
\providecommand{\BIBdecl}{\relax}
\BIBdecl

\bibitem{iannella:2022:obstructive}
G.~Iannella, G.~Magliulo, A.~Greco, M.~De~Vincentiis, M.~Ralli, A.~Maniaci, and
  C.~Vicini, ``Obstructive sleep apnea syndrome: from symptoms to treatment,''
  \emph{International Journal of Environmental Research and Public Health},
  vol.~19, no.~4, p. 2459, 2022.

\bibitem{zhao:2023:association}
Z.~Zhao, Y.~Gao, J.~Lin, R.~Xu, Z.~He, L.~Zhao, and L.~Liu, ``Association of
  depression with long-term cardiovascular risks in older patients with
  obstructive sleep apnea,'' \emph{Nature and Science of Sleep}, pp.
  1033--1043, 2023.

\bibitem{bousquet:2007:global}
J.~Bousquet, R.~Dahl, and N.~Khaltaev, ``Global alliance against chronic
  respiratory diseases,'' \emph{European Respiratory Journal}, vol.~29, no.~2,
  pp. 233--239, 2007.

\bibitem{strollo:1996:obstructive}
P.~J. Strollo~Jr and R.~M. Rogers, ``Obstructive sleep apnea,'' \emph{New
  England Journal of Medicine}, vol. 334, no.~2, pp. 99--104, 1996.

\bibitem{thornton:2012:aasm}
A.~T. Thornton, P.~Singh, W.~R. Ruehland, and P.~D. Rochford, ``Aasm criteria
  for scoring respiratory events: interaction between apnea sensor and hypopnea
  definition,'' \emph{Sleep}, vol.~35, no.~3, pp. 425--432, 2012.

\bibitem{pordzik:2024:waiting}
J.~Pordzik, C.~Seifen, K.~Ludwig, C.~Ruckes, T.~Huppertz, C.~Matthias, and
  H.~Gouveris, ``Waiting for in-lab polysomnography may unnecessarily prolong
  treatment start in patients with moderate or severe osa at home sleep apnea
  testing,'' \emph{Nature and Science of Sleep}, pp. 1881--1889, 2024.

\bibitem{tan:2014:overnight}
H.-L. Tan, D.~Gozal, H.~M. Ramirez, H.~P.~R. Bandla, and L.~Kheirandish-Gozal,
  ``Overnight polysomnography versus respiratory polygraphy in the diagnosis of
  pediatric obstructive sleep apnea,'' \emph{Sleep}, vol.~37, no.~2, pp.
  255--260, 2014.

\bibitem{kapur:2017:clinical}
V.~K. Kapur, D.~H. Auckley, S.~Chowdhuri, D.~C. Kuhlmann, R.~Mehra, K.~Ramar,
  and C.~G. Harrod, ``Clinical practice guideline for diagnostic testing for
  adult obstructive sleep apnea: an american academy of sleep medicine clinical
  practice guideline,'' \emph{Journal of clinical sleep medicine}, vol.~13,
  no.~3, pp. 479--504, 2017.

\bibitem{epstein:2024:use}
S.~Epstein and K.~Eng, ``Use of home sleep apnea testing to investigate
  persistent hypoxia in the remote setting,'' in \emph{SLEEP}, vol.~47.\hskip
  1em plus 0.5em minus 0.4em\relax Oxford University Press, May 2024.

\bibitem{cao:2024:automatic}
S.~Cao, I.~Rosenzweig, F.~Bilotta, H.~Jiang, and M.~Xia, ``Automatic detection
  of obstructive sleep apnea based on speech or snoring sounds: a narrative
  review,'' \emph{Journal of Thoracic Disease}, vol.~16, no.~4, pp. 2654--2667,
  2024.

\bibitem{kim:2025:sound}
S.~Kim, E.~Cho, H.~Kyung, J.~Hong, D.~Lee, K.~Y. Jung, and J.~W. Kim,
  ``Sound-based ai model for estimating apnea-hypopnea index in sleep apnea
  detection,'' in \emph{SLEEP}, vol.~48.\hskip 1em plus 0.5em minus 0.4em\relax
  Oxford University Press, May 2025, pp. A601--A602.

\bibitem{yaffe:2011:sleep}
K.~Yaffe, A.~M. Laffan, S.~L. Harrison, S.~Redline, A.~P. Spira, K.~E. Ensrud,
  S.~Ancoli-Israel, and K.~L. Stone, ``Sleep-disordered breathing, hypoxia, and
  risk of mild cognitive impairment and dementia in older women,'' \emph{JAMA},
  vol. 306, no.~6, pp. 613--619, 2011.

\bibitem{bubu:2020:obstructive}
O.~M. Bubu, A.~G. Andrade, O.~Q. Umasabor-Bubu, M.~M. Hogan, A.~D. Turner,
  M.~J. de~Leon, G.~Ogedegbe, G.~Jean-Louis, M.~L. Jackson, A.~W. Varga, and
  R.~S. Osorio, ``Obstructive sleep apnea, cognition and {Alzheimer's} disease:
  A systematic review integrating three decades of multidisciplinary
  research,'' \emph{Sleep Medicine Reviews}, vol.~50, p. 101250, 2020.

\bibitem{phan:2022:sleeptransformer}
H.~Phan, K.~Mikkelsen, O.~Y. Ch{\'e}n, P.~Koch, A.~Mertins, and M.~De~Vos,
  ``{SleepTransformer}: Automatic sleep staging with interpretability and
  uncertainty quantification,'' \emph{IEEE Transactions on Biomedical
  Engineering}, vol.~69, no.~8, pp. 2456--2467, 2022.

\bibitem{park2025machine}
C.~Park, J.~I. Byun, S.~H. Choi, and W.~C. Shin, ``Machine learning classifier
  solving the problem of sleep stage imbalance between overnight sleep,''
  \emph{Biomedical Engineering Letters}, vol.~15, no.~3, pp. 513--523, 2025.

\bibitem{Park2025}
\BIBentryALTinterwordspacing
C.~Park and C.~Kim, ``A novel chain-of-thought reasoning approach for
  alzheimer's disease detection using large language and vision-language
  models,'' \emph{IEEE Transactions on Neural Systems and Rehabilitation
  Engineering}, vol.~33, pp. 4386--4395, 2025. [Online]. Available:
  \url{http://dx.doi.org/10.1109/TNSRE.2025.3624762}
\BIBentrySTDinterwordspacing

\bibitem{park25d_interspeech}
C.~Park, A.~S.~G. Choi, S.~Cho, and C.~Kim, ``{Reasoning-Based Approach with
  Chain-of-Thought for Alzheimer's Detection Using Speech and Large Language
  Models},'' in \emph{{Interspeech 2025}}, 2025, pp. 2185--2189.

\bibitem{gonzalez:2024:audio}
F.~D. González~Martínez, J.~de~la Torre~Cruz, J.~J. Carabias~Orti, F.~J.
  Canadas~Quesada, A.~A. Salvador~Navarro, J.~Ranilla, and L.~Lamrini H~Laarbi,
  ``Audio-polygraphy dataset for sleep apnea analysis (apsaa),''
  \emph{Audio-Polygraphy Dataset for Sleep Apnea Analysis (APSAA)}, 2024.

\bibitem{cakir:2017:convolutional}
E.~Cakir, G.~Parascandolo, T.~Heittola, H.~Huttunen, and T.~Virtanen,
  ``Convolutional recurrent neural networks for polyphonic sound event
  detection,'' \emph{IEEE/ACM Transactions on Audio, Speech, and Language
  Processing}, vol.~25, no.~6, pp. 1291--1303, 2017.

\bibitem{gong:2021:ast}
Y.~Gong, Y.-A. Chung, and J.~Glass, ``Ast: Audio spectrogram transformer,''
  \emph{arXiv preprint arXiv:2104.01778}, 2021.

\bibitem{naman:2025:fast}
A.~Naman and G.~Zhang, ``Fast: Fast audio spectrogram transformer,'' in
  \emph{ICASSP 2025-2025 IEEE International Conference on Acoustics, Speech and
  Signal Processing (ICASSP)}.\hskip 1em plus 0.5em minus 0.4em\relax IEEE,
  Apr. 2025, pp. 1--5.

\bibitem{liu:2023:cat}
X.~Liu, H.~Lu, J.~Yuan, and X.~Li, ``Cat: Causal audio transformer for audio
  classification,'' in \emph{ICASSP 2023-2023 IEEE International Conference on
  Acoustics, Speech and Signal Processing (ICASSP)}.\hskip 1em plus 0.5em minus
  0.4em\relax IEEE, Jun. 2023, pp. 1--5.

\bibitem{gemmeke:2017:audio}
J.~F. Gemmeke, D.~P. Ellis, D.~Freedman, A.~Jansen, W.~Lawrence, R.~C. Moore,
  M.~Plakal, and M.~Ritter, ``Audio set: An ontology and human-labeled dataset
  for audio events,'' in \emph{ICASSP 2017-2017 IEEE International Conference
  on Acoustics, Speech and Signal Processing (ICASSP)}.\hskip 1em plus 0.5em
  minus 0.4em\relax New Orleans, LA: IEEE, Mar. 2017, pp. 776--780.

\bibitem{piczak:2015:esc}
K.~J. Piczak, ``Esc: Dataset for environmental sound classification,'' in
  \emph{Proceedings of the 23rd ACM international conference on Multimedia},
  Oct. 2015, pp. 1015--1018.

\bibitem{warden:2018:speech}
\BIBentryALTinterwordspacing
P.~Warden, ``Speech commands: A dataset for limited-vocabulary speech
  recognition,'' \emph{arXiv preprint arXiv:1804.03209}, Apr. 2018. [Online].
  Available: \url{https://arxiv.org/abs/1804.03209}
\BIBentrySTDinterwordspacing

\bibitem{wolf-monheim:2024:spectral}
\BIBentryALTinterwordspacing
F.~Wolf-Monheim, ``Spectral and rhythm features for audio classification with
  deep convolutional neural networks,'' \emph{arXiv preprint arXiv:2410.06927},
  Oct. 2024. [Online]. Available: \url{https://arxiv.org/abs/2410.06927}
\BIBentrySTDinterwordspacing

\bibitem{turab:2022:investigating}
\BIBentryALTinterwordspacing
M.~Turab, T.~Kumar, M.~Bendechache, and T.~Saber, ``Investigating multi-feature
  selection and ensembling for audio classification,'' \emph{arXiv preprint
  arXiv:2206.07511}, 2022. [Online]. Available:
  \url{https://arxiv.org/abs/2206.07511}
\BIBentrySTDinterwordspacing

\bibitem{deshmukh:2021:improving}
\BIBentryALTinterwordspacing
S.~Deshmukh, B.~Raj, and R.~Singh, ``Improving weakly supervised sound event
  detection with self-supervised auxiliary tasks,'' \emph{arXiv preprint
  arXiv:2106.06858}, 2021. [Online]. Available:
  \url{https://arxiv.org/abs/2106.06858}
\BIBentrySTDinterwordspacing

\bibitem{xin:2023:background}
Y.~Xin, D.~Yang, and Y.~Zou, ``Background-aware modeling for weakly supervised
  sound event detection,'' in \emph{Proc. INTERSPEECH}, vol. 2023, Jan. 2023,
  pp. 1199--1203.

\bibitem{shah:2018:closer}
\BIBentryALTinterwordspacing
A.~Shah, A.~Kumar, A.~G. Hauptmann, and B.~Raj, ``A closer look at weak label
  learning for audio events,'' \emph{arXiv preprint arXiv:1804.09288}, 2018.
  [Online]. Available: \url{https://arxiv.org/abs/1804.09288}
\BIBentrySTDinterwordspacing

\bibitem{miyazaki:2020:weakly}
K.~Miyazaki, T.~Komatsu, T.~Hayashi, S.~Watanabe, T.~Toda, and K.~Takeda,
  ``Weakly-supervised sound event detection with self-attention,'' in
  \emph{ICASSP 2020-2020 IEEE International Conference on Acoustics, Speech and
  Signal Processing (ICASSP)}.\hskip 1em plus 0.5em minus 0.4em\relax IEEE, May
  2020, pp. 66--70.

\bibitem{nguyen:2023:improving}
B.~Nguyen, S.~Uhlich, and F.~Cardinaux, ``Improving self-supervised learning
  for audio representations by feature diversity and decorrelation,'' in
  \emph{Icassp 2023-2023 ieee international conference on acoustics, speech and
  signal processing (icassp)}.\hskip 1em plus 0.5em minus 0.4em\relax IEEE,
  Jun. 2023, pp. 1--5.

\bibitem{yadav:2025:axlstms}
S.~Yadav, S.~Theodoridis, and Z.-H. Tan, ``Axlstms: learning self-supervised
  audio representations with xlstms,'' in \emph{Interspeech}, vol. 2025, 2025,
  pp. 3524--3528.

\bibitem{farhang:2013:adaptive}
B.~Farhang-Boroujeny, \emph{Adaptive Filters: Theory and Applications}.\hskip
  1em plus 0.5em minus 0.4em\relax John Wiley \& Sons, 2013.

\bibitem{lan:2023:variational}
H.~Lan, J.~Hu, Z.~Wang, and Q.~Cheng, ``Variational nonlinear kalman filtering
  with unknown process noise covariance,'' \emph{IEEE Transactions on Aerospace
  and Electronic Systems}, vol.~59, no.~6, pp. 9177--9190, 2023.

\bibitem{wen:2024:adaptive}
R.~Wen, H.~McCoy, D.~Tench, G.~Tagliavini, M.~A. Bender, A.~Conway, and
  P.~Pandey, ``Adaptive quotient filters,'' \emph{Proceedings of the ACM on
  Management of Data}, vol.~2, no.~4, pp. 1--28, 2024.

\bibitem{comminiello:2018:adaptive}
D.~Comminiello and J.~C. Príncipe, Eds., \emph{Adaptive Learning Methods for
  Nonlinear System Modeling}.\hskip 1em plus 0.5em minus 0.4em\relax
  Butterworth-Heinemann, 2018.

\bibitem{vaseghi:2008:advanced}
S.~V. Vaseghi, \emph{Advanced Digital Signal Processing and Noise
  Reduction}.\hskip 1em plus 0.5em minus 0.4em\relax John Wiley \& Sons, 2008.

\bibitem{sathesh:2015:application}
K.~Sathesh and N.~J.~R. Muniraj, ``Application of adaptive line enhancer with
  lms to separate heart sound signal from lung sound signal at real time,''
  \emph{Research Journal of Applied Sciences, Engineering and Technology},
  vol.~9, no.~6, pp. 448--453, 2015.

\bibitem{alipoor:2017:utilizing}
G.~Alipoor, ``Utilizing kernel adaptive filters for speech enhancement within
  the ale framework,'' \emph{Iranian Journal of Electrical \& Electronic
  Engineering}, vol.~13, no.~4, p. 303, 2017.

\bibitem{nakanishi:2013:speech}
I.~Nakanishi, H.~Namba, and S.~Li, ``Speech enhancement based on frequency
  domain ale with adaptive de-correlation parameters,'' \emph{International
  Journal of Computer Theory and Engineering}, vol.~5, no.~2, pp. 292--297,
  2013.

\bibitem{atkins:2021:adaptive}
A.~Atkins, I.~Cohen, and J.~Benesty, ``Adaptive line enhancer for nonstationary
  harmonic noise reduction,'' \emph{Computer Speech \& Language}, vol.~70, p.
  101245, 2021.

\bibitem{haykin:2002:adaptive}
S.~S. Haykin, \emph{Adaptive Filter Theory}.\hskip 1em plus 0.5em minus
  0.4em\relax Pearson Education India, 2002.

\bibitem{park:2014:noise}
M.~J. Park, J.~H. Yoo, B.~W. Cho, K.~T. Kim, W.~C. Jeong, and M.~Ha, ``Noise in
  hospital rooms and sleep disturbance in hospitalized medical patients,''
  \emph{Environmental Health and Toxicology}, vol.~29, p. e2014006, 2014.

\bibitem{siebein:2025:soundscape}
K.~Siebein, G.~Siebein, L.~Lorusso, D.~Rittenbacher, T.~Orozco, J.~Miller,
  R.~Skelton, and I.~Jones, ``The soundscape of an icu,'' \emph{The Journal of
  the Acoustical Society of America}, vol. 153, no. 3\_supplement, p. A294,
  2025.

\bibitem{wang:2017:methods}
C.~Wang and J.~Peng, ``The methods of acoustical analysis of snoring for the
  diagnosis of osahs,'' \emph{Journal of Sleep Medicine and Disorders}, vol.~4,
  no.~2, p. 1080, 2017.

\bibitem{kowallik:2001:breath}
P.~Kowallik, I.~Jacobi, A.~Jirmann, M.~Meesmann, M.~Schmidt, and H.~Wirtz,
  ``Breath-to-breath variability correlates with apnea-hypopnea index in
  obstructive sleep apnea,'' \emph{Chest}, vol. 119, no.~2, pp. 451--459, 2001.

\bibitem{hongzhang:2013:time}
Y.~Hong~Zhang, Q.~Lu~Li, J.~Wu, and J.~Li, ``The time-frequency characters
  analysis of the snoring signal,'' in \emph{2013 International Conference on
  Advanced Computer Science and Electronics Information (ICACSEI 2013)}.\hskip
  1em plus 0.5em minus 0.4em\relax Atlantis Press, 2013, pp. 83--86.

\bibitem{mikami:2015:detecting}
T.~Mikami, S.~Ueki, H.~Takahashi, and K.~Yonezawa, ``Detecting nonlinear
  acoustic properties of snoring sounds using hilbert-huang transform,'' in
  \emph{Proceedings of the International Conference on Bio-inspired Systems and
  Signal Processing (BIOSTEC)}, vol.~2.\hskip 1em plus 0.5em minus 0.4em\relax
  SCITEPRESS, 2015, pp. 306--311.

\bibitem{gulati:2020:conformer}
A.~Gulati, J.~Qin, C.-C. Chiu, N.~Parmar, Y.~Zhang, J.~Yu, W.~Han, S.~Wang,
  Z.~Zhang, Y.~Wu, and R.~Pang, ``Conformer: Convolution-augmented transformer
  for speech recognition,'' in \emph{Proceedings of the 21st Annual Conference
  of the International Speech Communication Association (Interspeech 2020)},
  2020, pp. 5036--5040.

\bibitem{loshchilov:2017:decoupled}
I.~Loshchilov and F.~Hutter, ``Decoupled weight decay regularization,''
  \emph{arXiv preprint arXiv:1711.05101}, 2017.

\bibitem{smith:2019:super}
L.~N. Smith and N.~Topin, ``Super-convergence: Very fast training of neural
  networks using large learning rates,'' in \emph{Proceedings of SPIE 11006,
  Artificial Intelligence and Machine Learning for Multi-Domain Operations
  Applications}, vol. 11006.\hskip 1em plus 0.5em minus 0.4em\relax
  International Society for Optics and Photonics, 2019, pp. 369--386.

\bibitem{gonzalez:2025:polygraph}
F.~D. Gonzalez-Martinez, J.~De~La Torre-Cruz, J.~J. Carabias-Orti, F.~J.
  Canadas-Quesada, A.~A. Salvador-Navarro, J.~Ranilla, and L.~L.~H. Laarbi,
  ``Polygraph and audio synchronization applied to apnea event analysis based
  on non-negative matrix factorization,'' \emph{EURASIP Journal on Audio,
  Speech, and Music Processing}, vol. 2025, no.~1, p.~24, 2025.

\bibitem{chang:2023:international}
J.~L. Chang, A.~N. Goldberg, J.~A. Alt, A.~Mohammed, L.~Ashbrook, D.~Auckley,
  ..., and I.~M. Rosen, ``International consensus statement on obstructive
  sleep apnea,'' in \emph{International forum of allergy \& rhinology},
  vol.~13, no.~7.\hskip 1em plus 0.5em minus 0.4em\relax Wiley, Jul. 2023, pp.
  1061--1482.

\bibitem{kim:2022:definition}
S.~G. Kim, S.~W. Cho, and J.~W. Kim, ``Definition of the snoring episode index
  based on the analyses of snoring parameters and the apnea hypopnea index,''
  \emph{Scientific Reports}, vol.~12, no.~1, p. 6761, 2022.

\bibitem{gould:1988:sleep}
G.~A. Gould, K.~F. Whyte, G.~B. Rhind, M.~A.~A. Airlie, J.~R. Catterall, C.~M.
  Shapiro, and N.~J. Douglas, ``The sleep hypopnea syndrome,'' \emph{American
  Review of Respiratory Disease}, vol. 137, no.~4, pp. 895--898, 1988.

\bibitem{mcnicholas:2008:diagnosis}
W.~T. McNicholas, ``Diagnosis of obstructive sleep apnea in adults,''
  \emph{Proceedings of the American Thoracic Society}, vol.~5, no.~2, pp.
  154--160, 2008.

\bibitem{bilen:2020:framework}
C.~Bilen, G.~Ferroni, F.~Tuveri, J.~Azcarreta, and S.~Krstulovic, ``A framework
  for the robust evaluation of sound event detection,'' in \emph{ICASSP
  2020-2020 IEEE International Conference on Acoustics, Speech and Signal
  Processing (ICASSP)}.\hskip 1em plus 0.5em minus 0.4em\relax IEEE, May 2020,
  pp. 61--65.

\end{thebibliography}

\end{document}